\documentclass[pra,superscriptaddress,showpacs,twocolumn]{revtex4}
\usepackage{amssymb}
\usepackage{graphicx}
\usepackage{amsmath}
\usepackage{ulem}
\newcommand{\subsetq}{\ensuremath%
        {\underset{\mbox{\sout{\tiny{\,?\,}}}}{\subset}}}
\newcommand{\supsetq}{\ensuremath%
        {\underset{\mbox{\sout{\tiny{\,?\,}}}}{\supset}}}

\def\ra{\ensuremath{\rightarrow}}
\def\da{\ensuremath{\downarrow}}
\newtheorem{thm}{Theorem}

\newtheorem{lem}[thm]{Lemma}

\newcommand{\proky}[1]{{P({#1}\otimes c\ra y\otimes
c)}}

\begin{document}

\title{Relation between catalyst-assisted transformation and multiple-copy transformation for bipartite pure states}
\author{Yuan Feng}
\email{feng-y@tsinghua.edu.cn}
\author{Runyao Duan}
\email{dry02@mails.tsinghua.edu.cn}
\author{Mingsheng Ying}
\email{yingmsh@tsinghua.edu.cn}

\address{State Key Laboratory of
Intelligent Technology and Systems, Department of Computer Science
and Technology Tsinghua University, Beijing, China, 100084}
\date{\today}
\begin{abstract}
We show that in some cases, catalyst-assisted entanglement
transformation cannot be implemented by multiple-copy transformation
for pure states. This fact, together with the result we obtained in
[R. Y. Duan, Y. Feng, X. Li, and M. S. Ying, Phys. Rev. A 71, 042319
(2005)] that the latter can be completely implemented by the former,
indicates that catalyst-assisted transformation is strictly more
powerful than multiple-copy transformation. For purely probabilistic
setting we find, however, these two kinds of transformations are
geometrically equivalent in the sense that the sets of pure states
which can be converted into a given pure state with maximal
probabilities not less than a given value have the same closure, no
matter catalyst-assisted transformation or multiple-copy
transformation is used.
\end{abstract}
 \maketitle

\section{Introduction}
Quantum entanglement, which is essential in quantum information
processing such as quantum cryptography \cite{BB84}, quantum
superdense coding \cite{BW92} and quantum teleportation
\cite{BBC+93}, has been extensively studied. One fruitful research
direction on quantum entanglement is to discuss the possibility of
transforming a bipartite entangled pure state into another one
allowing only local operations on the separate subsystems
respectively and classical communication between them (or LOCC for
short). The asymptotic case when arbitrarily large number of copies
are provided is considered by Bennett and his collaborators
\cite{BBPS96}. While in deterministic and finite manner, the first
and significant step was made by Nielsen \cite{NI99} who discovered
the connection between the theory of majorization in linear algebra
\cite{MO79} and entanglement transformation. Nielsen proved that a
bipartite entangled pure state $|\psi_1\rangle$ can be transformed
into another bipartite entangled pure state $|\psi_2\rangle$ by LOCC
if and only if $\lambda_{\psi_1}\prec\lambda_{\psi_2}$, where the
probability vectors $\lambda_{\psi_1}$ and $\lambda_{\psi_2}$ denote
the Schmidt coefficient vectors of $|\psi_1\rangle$ and
$|\psi_2\rangle$, respectively. Here the symbol $\prec$ stands for
`majorization relation'. Generally, an $n$-dimensional real vector
$x$ is said to be majorized by another $n$-dimensional real vector
$y$, denoted by $x\prec y$, if the following relations hold:
\begin{equation}
\sum_{i=1}^l x^\downarrow_i \leq \sum_{i=1}^l y^\downarrow_i {\rm \
\ \ for\ any\ \ \ }1\leq l\leq n,
\end{equation} with equality holding when $l=n$, where $x^\downarrow$
denotes the vector obtained by rearranging the components of $x$ in
nonincreasing order.

Nielsen's theorem gives a necessary and sufficient condition when
two entangled pure states are comparable in the sense that one can
be transformed into another by LOCC. There exist, however,
incomparable states such that any one cannot be transformed into
another only using LOCC. To treat the case of transformations
between incomparable states, Vidal \cite{Vi99} generalized Nielsen's
work by allowing probabilistic transformations. He found that the
maximal probability of transforming $|\psi_1\rangle$ into
$|\psi_2\rangle$ by LOCC can be calculated by
\begin{equation}\label{eq:Vidal}
P(|\psi_1\rangle \rightarrow |\psi_2\rangle)=\min_{1\leq l\leq n}
\frac{E_l(\lambda_{\psi_1})}{E_l(\lambda_{\psi_2})},
\end{equation}
where $E_l(x)$ denotes $\sum_{i=l}^n x^{\downarrow}_i$.

In Ref.\cite{JP99}, Jonathan and Plenio discovered a very surprising
phenomenon that sometimes an entangled state can enable otherwise
impossible entanglement transformations without being consumed at
all. A simple but well known example is $|\psi_1\rangle\nrightarrow
|\psi_2\rangle$ but $|\psi_1\rangle\otimes|\phi\rangle \rightarrow
|\psi_2\rangle\otimes|\phi\rangle$, where
$|\psi_1\rangle=\sqrt{0.4}|00\rangle+\sqrt{0.4}|11\rangle+\sqrt{0.1}|22\rangle+\sqrt{0.1}|33\rangle,$
$|\psi_2\rangle=\sqrt{0.5}|00\rangle+\sqrt{0.25}|11\rangle+\sqrt{0.25}|22\rangle,$
and $|\phi\rangle=\sqrt{0.6}|44\rangle+\sqrt{0.4}|55\rangle.$ The
role of the state $|\phi\rangle$ is just like a catalyst in a
chemical process. Daftuar and Klimesh \cite{DK01} examined
catalyst-assisted entanglement transformation and derived some
interesting results. In \cite{FD05a}, we investigated
catalyst-assisted transformation in probabilistic setting. A
necessary and sufficient condition was presented under which there
exist partial catalysts that can increase the maximal transforming
probability of a given entanglement transformation. The mathematical
structure of catalyst-assisted probabilistic transformation was also
carefully investigated.

Another interesting phenomenon of entanglement transformation was
noticed by Bandyopadhyay $et$ $al$. \cite{SRS02}. In some occasions,
increasing the number of copies of the original state can also help
entanglement transformations. Take the above example. Instead of
introducing a catalyst state $|\phi\rangle$, providing 3 copies of
$|\psi_{1}\rangle$ is also sufficient to transform these copies
together into the same number of $|\psi_{2}\rangle$. A question
naturally arises here is: what is the relation between
catalyst-assisted entanglement transformation and multiple-copy
transformation? In \cite{DF05b}, we found that multiple-copy
entanglement transformation can be completely implemented by
catalyst-assisted one. Furthermore, the mixing of these two has also
the same power as pure catalyst-assisted transformation. In other
words, any transformation which can be realized collectively on
multiple copies and with the aid of a catalyst can be exactly
implemented by only providing some appropriate catalyst. Later on,
we proved that these two kinds of transformations are asymptotically
equivalent in the sense that they can simulate each other's ability
to implement a desired transformation with the same optimal success
probability, when the dimension of catalysts and the number of
copies provided tend to infinity \cite{DF05a}.

The contribution of the current paper is twofold. First, we show
that in some cases catalyst-assisted entanglement transformation is
strictly more powerful than multiple-copy one by deriving a
sufficient condition when the former cannot be implemented by the
latter. Second, for purely probabilistic setting we find, however,
these two kinds of transformations are geometrically equivalent.
That is, no matter catalyst-assisted transformations or
multiple-copy transformations are used, the sets of quantum states
that can be converted into a given state with maximal probabilities
not less than a given value have the same closure. It is worth
noting that the geometrical equivalence between these two kinds of
transformations proved in the current paper is different from the
asymptotical equivalence shown in \cite{DF05a}. We will elaborate
the difference at the end of Section III after necessary notations
have been introduced.

For simplicity, in what follows we denote a bipartite pure state by
the probability vector of its Schmidt coefficients. This will not
cause any confusion because it is well known that the fundamental
properties of a bipartite pure state under LOCC are completely
determined by its Schmidt coefficients. Therefore, from now on, we
consider only probability vectors (sometimes we even omit the
normalization of a nonnegative vector to be a probability one)
instead of quantum states and always identify a probability vector
with the bipartite pure state represented by it.

\section{Deterministic case}

In this section, we study the relation between catalyst-assisted
transformation and multiple-copy transformation in deterministic
case. First, we introduce some notations.

Denote by $V^n$ the set of all $n$-dimensional nonnegative vectors
and let $x,y,\cdots$ range over $V^n$. Let
\begin{equation}S(y)=\{x\in V^n\ |\ x\prec y\}\end{equation}
be the set of states that can be transformed into $y$ by LOCC
directly,
\begin{equation}T(y)=\{x\in V^n\ |\ \exists \mbox{ probability vector } c,\ x\otimes c\prec y\otimes c\}\end{equation} be the set of states
that can be transformed into $y$ by LOCC with the aid of some
catalyst, and
\begin{equation}M(y)=\{x\in V^n\ |\ \exists \mbox{ integer }k\ \geq 1,\ x^{\otimes{k}}\prec y^{{\otimes{k}}}\}\end{equation} the set of states which,
when some appropriate number of copies are provided, can be
transformed into the same number of $y$ by LOCC.

\begin{lem} Suppose $x\in T(y)$ and $x'\in T(y')$. Then
$\bar{x}\in T(\bar{y})$ where $\bar{x}= x\oplus x'$ and $\bar{y}=
y\oplus y'$.
\end{lem}
{\it Proof.} By definition, $x\in T(y)$ and $x'\in T(y')$ imply that
there exist $c$ and $c'$ such that $x\otimes c\prec y\otimes c$ and
$x'\otimes c'\prec y'\otimes c'$. It can be easily checked that the
vector $c\otimes c'$ serves as a catalyst for the transformation
from $\bar{x}$ to $\bar{y}$, that is, $\bar{x}\otimes c \otimes
c'\prec \bar{y}\otimes c\otimes c'$. Thus $\bar{x}\in
T(\bar{y})$.\hfill$\Box$

\vspace{1em}

The following lemma, important in its own right, is a powerful tool
which gives us a sufficient condition on $x$ and $y$ such that they
are incomparable in $any$ multiple-copy transformations. In other
words, $any$ number of $x$ cannot be collectively transformed into
the same number of $y$ using LOCC.

\begin{lem} Suppose $x$ and $y$ are two nonincreasingly arranged
$n$-dimensional probability vectors, $x_1=y_1$ but $x\nprec y$. Let
\begin{equation}d=\min\{l : 1\leq l \leq n,\ \sum_{i=1}^l x_i > \sum_{i=1}^l
y_i\}.\end{equation} Denote by $t_1$ the number of components in $x$
which are equal to $x_1$, while $t_2$ the number of components in
$y$ which are equal to $y_1$. If $t_1=t_2=t$ and
\begin{equation}\label{eq:ass0}
x_1x_d\geq x_{t+1}^2 \mbox{\ \ \ \ and\ \ \ \ } y_1y_d\geq
y_{t+1}^2,
\end{equation}
then $x\not\in M(y)$.
\end{lem}
{\it Proof.} First, it is obvious that $1\leq t< d<n$. From the
assumption Eq.(\ref{eq:ass0}), we have for any integer $k\geq 1$,
the components of $x^{\otimes k}$ and $y^{\otimes k}$ can be
arranged nonincreasingly as follows
\begin{equation}
\begin{array}{rcl}
(x^{\otimes k})^\downarrow&=& x_1^k\oplus (x_1^{k-1}x_2)^{\oplus
k}\oplus\cdots\oplus (x_{t}^{k-1}x_{t-1})^{\oplus k}\oplus
x_t^k\\
 &&\oplus (x_1^{k-1}x_{t+1})^{\oplus k}\oplus\cdots\oplus
(x_{t}^{k-1}x_{t+1})^{\oplus
k}\oplus\cdots\\
&&\oplus(x_{1}^{k-1}x_{d})^{\oplus
k}\oplus\cdots\oplus(x_{t}^{k-1}x_{d})^{\oplus k}\oplus\cdots
\end{array}
\end{equation} and
\begin{equation}
\begin{array}{rcl}
(y^{\otimes k})^\downarrow&=& y_1^k\oplus (y_1^{k-1}y_2)^{\oplus
k}\oplus\cdots\oplus (y_{t}^{k-1}y_{t-1})^{\oplus k}\oplus
y_t^k\\
 &&\oplus (y_1^{k-1}y_{t+1})^{\oplus k}\oplus\cdots\oplus
(y_{t}^{k-1}y_{t+1})^{\oplus
k}\oplus\cdots\\
&&\oplus(y_{1}^{k-1}y_{d})^{\oplus
k}\oplus\cdots\oplus(y_{t}^{k-1}y_{d})^{\oplus k}\oplus\cdots
\end{array}
\end{equation}
where by $\alpha^{\oplus k}$ we denote the vector
\begin{equation}
\begin{array}{c}
\underbrace{(\alpha,\ \alpha,\ \cdots,\ \alpha)}.\\
k\mbox{\ times}
\end{array}
\end{equation}
Here we only write out explicitly the largest $t^k+kt(d-t)$
components of $x^{\otimes k}$ and $y^{\otimes k}$ since it is enough
for our argument. Notice that $x_1=\cdots =x_t>x_{t+1}$ and
$y_1=\cdots =y_t>y_{t+1}$. It is now easy to check that $x^{\otimes
k}\nprec y^{\otimes k}$ since when taking $l=t^k+kt(d-t)$, we have
\begin{eqnarray}
\sum\limits_{i=1}^l (x^{\otimes k})^{\downarrow}_i &=& (tx_1)^k +
ktx_1^{k-1}\sum\limits_{i=t+1}^d x_i \\
&>& (ty_1)^k + kty_1^{k-1}\sum\limits_{i=t+1}^d y_i\\
&=& \sum\limits_{i=1}^l (y^{\otimes k})^{\downarrow}_i.
\end{eqnarray}
So $x\not\in M(y)$ by the arbitrariness of $k$.\hfill$\Box$

 \vspace{1em}

If we take $x$ and $y$ as in Lemma 2 but $x_n=y_n$ instead of
$x_1=y_1$. Let
\begin{equation}d=\max\{l : 1\leq l \leq n,\ \sum_{i=l}^n x_i <
\sum_{i=l}^n
y_i\}.\end{equation} Denote by $t_1$ the number of components in $x$
which are equal to $x_n$ while $t_2$ the number of components in $y$
which are equal to $y_n$. If $t_1=t_2=t$ , $x_nx_d\leq x_{n-t}^2$,
and $y_ny_d\leq y_{n-t}^2$, then we can also deduce that $x\not\in
M(y)$.

 \vspace{1em}

Using the lemmas above, we can now prove that $T(y)\not =M(y)$ for
some probability vector $y$ by deriving a sufficient condition
under which $T(y)\not\subseteq M(y)$, as the following theorem
states.

\begin{thm}\label{thm:mrdeterm} Suppose $y$ is a nonincreasingly arranged $n$-dimensional
probability vector. Denote by $t$ and $m$ the numbers of components
which are equal to $y_1$ and which are equal to $y_n$, respectively.
Let $d$ be the minimal index of the components which are less than
$y_{t+1}$. That is,
\begin{equation}
y_1=\cdots=y_{t}>y_{t+1}=\cdots=y_{d-1}>y_{d},
\end{equation} and
\begin{equation}
y_{n-m}>y_{n-m+1}=\cdots=y_n.
\end{equation}
If $d<n-m$ and $y_1y_{d}\geq y_{t+1}^2$, then $T(y)\not\subseteq
M(y)$.
\end{thm}
{\it Proof.} Take a positive number $\epsilon$ such that
\begin{equation}
\epsilon< \min\{\frac{d-t-1}{d-t}(y_{d-1}-y_{d}), \
\frac{m}{m+1}(y_{n-m}-y_{n-m+1})\}.
\end{equation}
Define two $(n-t)$-dimensional nonnegative vectors $\bar{x}$ and
$\bar{y}$ as follows
\begin{equation}
\begin{array}{rcl}
\displaystyle\bar{x}&=&(y_{t+1}-\displaystyle\frac{\epsilon}{\triangle},
\ \cdots,\  y_{d-1}-\frac{\epsilon}{\triangle},\ y_{d}+\epsilon,\
y_{d+1},\ \cdots, \\ \\
&& y_{n-m-1},\ y_{n-m}-\epsilon,\ \displaystyle
y_{n-m+1}+\frac{\epsilon}{m}, \ \cdots,\ y_n+\frac{\epsilon}{m})
\end{array}
\end{equation} and
\begin{equation}\bar{y}= (y_{t+1},\ y_{t+2},\ \cdots,\ y_n).
\end{equation}
Here $\triangle=d-t-1$. It is easy to check that $\bar{x}$ and
$\bar{y}$ are both nonincreasingly arranged, and $\bar{x}\prec
\bar{y}$. Furthermore, $\bar{x}$ is in the interior of $T(\bar{y})$
by Lemma 1 in \cite{DK01} since $\bar{x}_1<\bar{y}_1$ and
$\bar{x}_{n-t}>\bar{y}_{n-t}$. So there exists a sufficiently small
but positive $\delta$ such that $\bar{x}'\in T(\bar{y})$ where
\begin{equation}
\begin{array}{l}
\displaystyle\bar{x}'^{\da}=(y_{t+1}-\displaystyle\frac{\epsilon}{\triangle},
\ \cdots,\  y_{d-1}-\frac{\epsilon}{\triangle},\
y_{d}+\epsilon+\delta,\
y_{d+1},\ \cdots, \\ \\
y_{n-m-1},\ y_{n-m}-\epsilon-\delta,\ \displaystyle
y_{n-m+1}+\frac{\epsilon}{m}, \ \cdots,\ y_n+\frac{\epsilon}{m}).
\end{array}
\end{equation}
Now define $x$ as the direct sum of the vectors $(y_1,\dots,y_{t})$
and $\bar{x}'^{\da}$, that is
\begin{equation}
x= (y_1,\cdots, y_{t}, \bar{x}'^{\da}).
\end{equation}
By Lemma 1 we have $x\in T(y)$. On the other hand,
\begin{equation}\sum_{i=1}^{d} x_i = \sum_{i=1}^{d} y_i + \delta> \sum_{i=1}^{d} y_i,\end{equation}
\begin{equation}
\sum_{i=1}^{l} x_i \leq \sum_{i=1}^{l} y_i {\ \ \ \rm for \ \ \ \
}1\leq l<d,\end{equation} and
\begin{equation}
\begin{array}{rcl}
x_1x_{d}&=&\displaystyle y_1(y_{d}+\epsilon+\delta)\\ \\
&>&y_1y_{d}\geq y_{t+1}^2\\
\\&>&\displaystyle (y_{t+1}-\frac{\epsilon}{\triangle})^2=x_{t+1}^2,
\end{array}
\end{equation}
so we have $x\not \in M(y)$ from Lemma 2. That completes our
proof.\hfill$\Box$

 \vspace{1em}

Suppose $t$ and $m$ denote the numbers of the components which are
equal to $y_n$ and which are equal to $y_1$, respectively. Let $d$
be the maximal index of the components which are greater than
$y_{n-t}$. If $d>m+1$ and $y_ny_{d}\leq y_{n-t}^2$ then we can also
deduce that $T(y)\not\subseteq M(y)$.

 \vspace{1em}
An interesting special case of Theorem \ref{thm:mrdeterm} is when
$n>4$, if $y_1>y_2>y_3>y_{n-1}>y_n$ and $y_1y_3\geq y_2^2$ then
$T(y)\not\subseteq M(y)$.

Theorem \ref{thm:mrdeterm} in fact gives us a method to construct a
vector $y$ for which $T(y)\not\subseteq M(y)$. To be more specific,
given a vector $\bar{y}$ such that $T(\bar{y})\neq S(\bar{y})$, we
can derive a desired $y$ by the following two steps. First, add a
sufficiently large component to $\bar{y}$ such that the conditions
presented in Theorem \ref{thm:mrdeterm} are satisfied for the new
vector (notice that from Theorem 6 of \cite{DK01}, when
$T(\bar{y})\neq S(\bar{y})$, the condition $d<n-m$ in Theorem
\ref{thm:mrdeterm} holds automatically); second, normalize the
vector to $y$ such that it is a probability vector. For example,
given $\bar{y}=(0.5,0.25,0.25,0)$, we can derive
$y=(3,0.5,0.25,0.25,0)/4$ and $T(y)\not\subseteq M(y)$. Furthermore,
since the proof of Theorem \ref{thm:mrdeterm} is constructive, the
states which can be transformed into $y$ by catalyst-assisted
transformation while cannot by multiple-copy transformation can also
be constructed.

We have proved that $T(y)\not = M(y)$ in some cases. Moreover,
witness vectors which are in $T(y)$ but not in $M(y)$ are also
constructed explicitly. It should be pointed out, however, that the
witness vectors we constructed lie on the boundary of $T(y)$ without
any exception, that is, they all satisfy the property that
$x^\downarrow_1 = y^\downarrow_1$ or $x^\downarrow_n =
y^\downarrow_n$. These witness vectors can be involved if we
consider the closure of $M(y)$ instead. In fact, we will see in the
following section that in probabilistic setting, the two sets
$M^{\lambda}(y)$ and $T^{\lambda}(y)$ defined in Eqs.(\ref{eq:ty})
and (\ref{eq:my}) have exactly the same closure for $0\leq \lambda<
1$. So the question remained is to show whether or not ${M(y)}$ and
${T(y)}$ also have the same closure.

\section{Probabilistic case}

We considered deterministic entanglement transformations in the
previous section. In this section, let us turn to examine
transformations with maximal probability strictly less than $1$.

Given a nonnegative number $\lambda <1$, let
\begin{equation}\label{eq:sy}S^{\lambda}(y)=\{x\in V^n\ |\ P(x\ra y)\geq \lambda \}\end{equation} be the
set of states that can be transformed into $y$ by LOCC with the
maximal probability not less than $\lambda$,
\begin{equation}\label{eq:ty}T^{\lambda}(y)=\{x\in V^n\ |\ \exists
c, P(x\otimes c\ra y\otimes c)\geq \lambda \}\end{equation} be the
set of states that can be transformed into $y$ by catalyst-assisted
LOCC with the maximal probability not less than $\lambda$, and
\begin{equation}\label{eq:my}M^{\lambda}(y)=\{x\in V^n\ |\ \exists k, P(x^{\otimes{k}}\ra y^{{\otimes{k}}})^{1/k} \geq \lambda\}\end{equation}
the set of states which, when some appropriate number of copies are
provided, can be transformed into the same number of $y$ by
multiple-copy LOCC with the maximal geometric average probability
not less than $\lambda$. We have proved in \cite{DF05a} that
$M^{\lambda}(y)\subseteq T^{\lambda}(y)$. In the following we
further show that the reverse is not always true. For this purpose,
two lemmas which corresponding to Lemma 1 and Lemma 2 in the
previous section are useful.

\begin{lem}
Suppose $x,y\in V^n$ and $z\in V^m$ are nonincreasingly arranged
nonnegative vectors. If $x\in T^{\lambda}(y)$ then $x'\oplus \lambda
z\in T^{\lambda}(y\oplus z)$, where $x'=(x_1',x_2,\dots,x_n)$ with
$x_1'=x_1+(1-\lambda)\sum_{i=1}^m z_i$.
\end{lem}
{\it Proof.} From $x\in T^{\lambda}(y)$, there exists $c\in V^r$
such that $P(x\otimes c\ra y\otimes c)\geq \lambda$. For any
arbitrarily $1\leq l\leq (n+m)r$, we have
\begin{equation}\label{eq:28}
\sum_{i=l}^{(n+m)r} \left((x'\oplus \lambda z)\otimes c\right)_i^\da
= \sum_{i=l_1}^{nr}(x'\otimes c)_i^\da +
\lambda\sum_{i=l_2}^{mr}(z\otimes c)_i^\da
\end{equation}
for some $1\leq l_1\leq nr$ and $1\leq l_2\leq mr$. On the other
hand, by definition
\begin{equation}
\sum_{i=l}^{(n+m)r} \left((y\oplus z)\otimes c\right)_i^\da \leq
\sum_{i=l_1}^{nr}(y\otimes c)_i^\da + \sum_{i=l_2}^{mr}(z\otimes
c)_i^\da.
\end{equation}
Notice that $x_i'\geq x_i$ for any $i=1,\dots,n$, and that from
$P(x\otimes c\ra y\otimes c)\geq \lambda$ we know
\begin{equation}\label{eq:30}
\sum_{i=l_1}^{nr}(x\otimes c)_i^\da \geq
\lambda\sum_{i=l_1}^{nr}(y\otimes c)_i^\da.
\end{equation}
It follows from Eqs.(\ref{eq:28})-(\ref{eq:30}) that
\begin{equation} \sum_{i=l}^{(n+m)r} \left((x'\oplus \lambda
z)\otimes c\right)_i^\da \geq \lambda\sum_{i=l}^{(n+m)r}
\left((y\oplus z)\otimes c\right)_i^\da,
\end{equation}
and $P((x'\oplus \lambda z)\otimes c\ra (y\oplus z)\otimes c)\geq
\lambda$ from the arbitrariness of $l$. So we have $x'\oplus \lambda
z\in T^{\lambda}(y\oplus z)$.
 \hfill $\Box$

\begin{lem} Suppose $x$ and $y$ are two nonincreasingly arranged
$n$-dimensional probability vectors, $x_n=\lambda y_n$ but $x\not\in
S^{\lambda}(y)$ for $\lambda\in (0,1)$. Let
\begin{equation}d=\max\{l : 1\leq l \leq n,\ \sum_{i=l}^n x_i <
\lambda \sum_{i=l}^n y_i\}.\end{equation} Denote by $t_1$ the number
of components in $x$ which are equal to $x_n$, while $t_2$ the
number of components in $y$ which are equal to $y_n$. If $t_1=t_2=t$
and
\begin{equation}
x_nx_d\leq x_{n-t}^2 \mbox{\ \ \ \ and\ \ \ \ } y_ny_d\leq
y_{n-t}^2,
\end{equation}
then $x\not\in M^{\lambda}(y)$.
\end{lem}
{\it Proof.} Similar to Lemma 2. But the last $t^k+kt(n-t-d)$
components of $(x^{\otimes k})^\da$ and $(y^{\otimes k})^\da$ are
considered for any $k$ at this time. \hfill $\Box$

\begin{thm} Let $\lambda\in (0,1)$. There exists $y\in V^n$ such that $T^{\lambda}(y)\nsubseteq
M^{\lambda}(y)$.
\end{thm}
{\it Proof.} The proof is similar to but more complicated than that
of Theorem \ref{thm:mrdeterm}. Beside, due to the asymmetry of roles
of the largest and the smallest components in determining the
maximal transforming probability presented in Eq.(\ref{eq:Vidal}),
components at the tail but not at the head of $y$ should be
examined. We outline the main steps of the proof here.

Let $y\in V^n$ such that
\begin{equation}
y_1=\cdots=y_{m}>y_{m+1},
\end{equation} and
\begin{equation}
y_{d}>y_{d+1}=\cdots=y_{n-t}>y_{n-t+1}=\cdots=y_n.
\end{equation}
where $d>m+1$ and $y_ny_{d}\leq y_{n-t}^2$. Take a positive number
$\epsilon$ such that
\begin{equation}
\epsilon< \min\{\frac{\lambda
m}{m+1}(y_{m}-y_{m+1}),\frac{\lambda\triangle}{\triangle+1}(y_{d}-y_{d+1})\},
\end{equation}
where $\triangle=n-t-d$. Define
\begin{equation}
\begin{array}{l}
\displaystyle\bar{x}=(\widetilde{y_1},\lambda
y_{2}-\displaystyle\frac{\epsilon}{m}, \ \cdots,\ \lambda
y_{m}-\frac{\epsilon}{m},\ \lambda y_{m+1}+\epsilon,\
\lambda y_{m+2}, \\ \\
 \ \cdots,\lambda y_{d-1},\ \lambda y_{d}-\epsilon,\ \displaystyle
\lambda y_{d+1}+\frac{\epsilon}{\triangle}, \ \cdots,\ \lambda
y_{n-t}+\frac{\epsilon}{\triangle})
\end{array}
\end{equation} and
\begin{equation}\bar{y}= (y_{1},\ y_{2},\ \cdots,\ y_{n-t}).
\end{equation}
Here $\widetilde{y_1}=y_1+(1-\lambda)\sum_{i=2}^{n-t}
y_i-\epsilon/m$. Then $\bar{x}$ and $\bar{y}$ are both
nonincreasingly arranged, and $\bar{x}\in S^{\lambda}(\bar{y})$.
Furthermore, $\bar{x}$ is in the interior of $T^{\lambda}(\bar{y})$
by Theorem 9 in \cite{FD05a} since
$\bar{x}_{n-t}>\lambda\bar{y}_{n-t}$. So there exists a sufficiently
small but positive $\delta$ such that $\bar{x}'\in
T^{\lambda}(\bar{y})$ where
\begin{equation}
\begin{array}{rcl}
\bar{x}'^\da&=&(\widetilde{y_1},\lambda
y_{2}-\displaystyle\frac{\epsilon}{m}, \ \cdots,\ \lambda
y_{m}-\frac{\epsilon}{m},\ \lambda y_{m+1}+\epsilon+\delta,\\ \\
&&\lambda y_{m+2}, \ \cdots,\lambda y_{d-1},\ \lambda
y_{d}-\epsilon-\delta,\ \displaystyle \lambda
y_{d+1}+\frac{\epsilon}{\triangle}, \\ \\
&& \cdots,\ \lambda
y_{n-t}+\displaystyle\frac{\epsilon}{\triangle}).
\end{array}
\end{equation}
Now define $x= (x', \lambda y_{n-t+1},\cdots, \lambda y_{n})$, where
$x'=(x'_1,\bar{x}'^\da_2,\dots,\bar{x}'^\da_{n-t})$ and
$x'_1=y_1+(1-\lambda)\sum_{i=2}^{n} y_i-\epsilon/m$. By Lemma 4 and
5 we can similarly prove that $x \in T^{\lambda}(y)$ but $x\not\in
M^{\lambda}(y)$. \hfill$\Box$

\vspace{1em}

To draw a clearer picture of the relation between catalyst-assisted
transformation and multiple-copy transformation in purely
probabilistic setting, we investigate the limit properties of
$T^{\lambda}(y)$ and $M^{\lambda}(y)$ about $\lambda$. Since
$T^{\lambda'}(y)\subseteq T^{\lambda}(y)$ for any
$\lambda'>\lambda$, we can define
\begin{equation}T^{\lambda-}(y)=\bigcap_{\lambda'<\lambda}
T^{\lambda'}(y),\ \ \ \ \ \
T^{\lambda+}(y)=\bigcup_{\lambda'>\lambda}
T^{\lambda'}(y)\end{equation} which denote respectively the left
limit and right limit of the set-valued function $T^{\lambda}(y)$ at
the point $\lambda$. Similar notions can be defined for
$M^{\lambda+}(y)$ and $M^{\lambda-}(y)$. It is direct from the
definition that
\begin{eqnarray}
T^{\lambda-}(y)&=&\{\ x\ |\ \sup_c P(x\otimes c\ra y\otimes
c)\geq \lambda \},\\
M^{\lambda-}(y)&=&\{\ x\ |\ \sup_k P(x^{\otimes{k}}\ra
y^{{\otimes{k}}})^{1/k}\geq \lambda \},
\end{eqnarray}
and we have shown in \cite{DF05a} that
$T^{\lambda-}(y)=M^{\lambda-}(y)$ for any $\lambda\in [0,1]$.

The following theorem tells us that generally, the function
$T^{\lambda}(y)$ is neither left continuous nor right continuous at
any point $\lambda\in (0,1)$, although it is `almost' right
continuous in the sense that the right limit at $\lambda$ shares the
same interior points with $T^{\lambda}(y)$.

\begin{thm}\label{lem:tplus} For any $y\in V^n$ and $0<\lambda<1$,

1) $T^{\lambda+}(y)$ is open while $T^{\lambda-}(y)$ is closed,

2) $T^{\lambda+}(y)\varsubsetneq T^{\lambda}(y)\subseteq
T^{\lambda-}(y)$, and when $y^{\downarrow}_n>0$, $T^{\lambda}(y)=
T^{\lambda-}(y)$ if and only if $y^{\downarrow}_2=y^{\downarrow}_n$,

3) $T^{\lambda}(y)^{\circ}= T^{\lambda+}(y)$.
\end{thm}
{\it Proof.} 1). For any $x\in T^{\lambda+}(y)$, there exist
$\mu>\lambda$ and $c$ such that $P(x\otimes c\ra y\otimes c)\geq
\mu$. Let $\nu$ be a number such that $\mu>\nu>\lambda$. From the
continuity of $\proky{x}$ about $x$ for fixed $c$ and $y$, we have
some $\epsilon>0$, such that for any $x'\in B(x,\epsilon)$,
\begin{equation}
|\proky{x'}-\proky{x}|<\mu-\nu.
\end{equation}
We then derive that $x'\in T^{\lambda+}(y)$ since
\begin{equation}
\begin{array}{rl}
\proky{x'}&\geq \proky{x}-A \\
\\
&> \mu-(\mu-\nu)=\nu
\end{array}
\end{equation}
where $A=|\proky{x'}-\proky{x}|$. That completes the proof that
$T^{\lambda+}(y)$ is open. To prove the closeness of
$T^{\lambda-}(y)$, we take any sequence $x_i\in T^{\lambda-}(y)$
such that $\lim_i x_i= x$. By definition, for any $x_i$ we have
$\sup_c \proky{x_i}\geq \lambda$. To realize the transformation from
$x\otimes c$ to $y\otimes c$, a possible protocol is first
transforming $x$ to $x_i$ and then transforming $x_i\otimes c$ to
$y\otimes c$. So we have the following relation
\begin{equation}
\proky{x}\geq P(x\ra x_i)\proky{x_i}.
\end{equation}
Thus
\begin{equation}
\begin{array}{l}
\displaystyle\sup_c \proky{x}\\ \\
\geq P(x\ra x_i)\displaystyle\sup_c\proky{x_i}\\ \\\geq \lambda
P(x\ra x_i).
\end{array}
\end{equation}
The desired result that $x\in T^{\lambda-}(y)$ follows from the
above equation by letting $i$ tend to infinity.

2) is obvious from 1) and the fact that $T^{\lambda}(y)$ is neither
closed nor open when $y^{\downarrow}_2>y^{\downarrow}_n$ (see the
first two lines of the proof of Theorem 11 of \cite{FD05a}). When
$y^{\downarrow}_2=y^{\downarrow}_n$, we have
$T^{\lambda}(y)=S^{\lambda}(y)$ (Theorem 10 of \cite{FD05a}), then
$T^{\lambda}(y)= T^{\lambda-}(y)$ follows from the continuity of
$S^{\lambda}(y)$ for any $0<\lambda<1$ (Theorem 3 of \cite{FD05a}).

Now we prove 3). The relation $T^{\lambda+}(y)\subseteq
T^{\lambda}(y)^{\circ}$ is easy from 1) and 2). To prove the reverse
relation, we take any $x\in T^{\lambda}(y)^{\circ}$. Then
$\proky{x}\geq \lambda$ for some $c$. There are two cases to
consider.

Case 1. $\proky{x}>\lambda$. In this case, we know immediately that
$x\in T^{\lambda'}(y)\subseteq T^{\lambda+}(y)$ for
$\lambda'=\proky{x}$.

Case 2. $\proky{x}=\lambda$. Since $x$ is an interior point of
$T^{\lambda}(y)$, from Theorem 9 of \cite{FD05a} we have
$x^{\downarrow}_n/y^{\downarrow}_n>\lambda$. Then
$\proky{x}<\min\{1,x^{\downarrow}_n/y^{\downarrow}_n\}$. By Theorem
2 of \cite{FD04}, there exists a catalyst $c'$ such that
\begin{equation}
P(x\otimes c\otimes c'\ra y\otimes c\otimes c')>\proky{x}=\lambda.
\end{equation}
So we also have $x\in T^{\lambda'}(y)\subseteq T^{\lambda+}(y)$ for
$\lambda'=P(x\otimes c\otimes c'\ra y\otimes c\otimes c')$. \hfill
$\Box$

\vspace{1em} Notice that we assume $y^\da_n>0$ in 2) of the above
theorem. When $y^\da_n=0$, it is not clear till now whether or not
the result still holds.

With similar techniques, we can prove a corresponding result of
Theorem \ref{lem:tplus} for probabilistic multiple-copy
transformation.
\begin{thm}\label{lem:Mplus} For any $y\in V^n$ and $0<\lambda<1$,

1) $M^{\lambda+}(y)$ is open while $M^{\lambda-}(y)$ is closed,

2) $M^{\lambda+}(y)\varsubsetneq M^{\lambda}(y)\subseteq
M^{\lambda-}(y)$, and when $y^{\downarrow}_n>0$, $M^{\lambda}(y)=
M^{\lambda-}(y)$ if and only if $y^{\downarrow}_2=y^{\downarrow}_n$,

3) $M^{\lambda}(y)^{\circ}= M^{\lambda+}(y)$.
\end{thm}
{\it Proof.} Similar to the proof of Theorem \ref{lem:tplus}. \hfill
$\Box$

\vspace{1em}

Now we can show our main result of this section. Rather
surprisingly, when the probability threshold $\lambda$ is strictly
less than 1, probabilistic catalyst-assisted transformation and
probabilistic multiple-copy transformation are geometrically
equivalent in the sense that the two sets $T^{\lambda}(y)$ and
$M^{\lambda}(y)$ in fact share the same interior points (or
equivalently, the same closure).

\begin{lem}\label{lem:TleqM} If $0\leq \lambda<\lambda'\leq 1$, then
$T^{\lambda'}(y)\subseteq M^{\lambda}(y)$.
\end{lem}
{\it Proof.} By definition, for any $x\in T^{\lambda'}(y)$, there
exists $c$ such that $P(x\otimes c\ra y\otimes c)\geq \lambda'$.
Then from Theorem 1 of \cite{DF05a}, we have
\begin{equation}
\sup_k P(x^{\otimes{k}}\ra y^{{\otimes{k}}})^{1/k}=\sup_c P(x\otimes
c\ra y\otimes c)\geq \lambda'>\lambda.
\end{equation}
Thus there exists $k_0$ such that $P(x^{\otimes{k_0}}\ra
y^{{\otimes{k_0}}})^{1/k_0}>\lambda$. So $x\in M^{\lambda}(y)$.
\hfill $\Box$

\begin{thm}
For any $y\in V^n$ and $0\leq \lambda <1$, we have
$\overline{T^{\lambda}(y)}=\overline{M^{\lambda}(y)}$.
\end{thm}
{\it Proof.} From Theorems \ref{lem:tplus} and \ref{lem:Mplus}, to
prove this theorem we need only show that $T^{\lambda+}(y)=
M^{\lambda+}(y)$, or $T^{\lambda+}(y)\subseteq M^{\lambda+}(y)$
since the reverse is obvious. For any $\lambda'>\lambda$, take $\mu$
such that $\lambda'>\mu>\lambda$. Then from Lemma \ref{lem:TleqM} we
have $T^{\lambda'}(y)\subseteq M^{\mu}(y)\subseteq M^{\lambda+}(y)$.
So $T^{\lambda+}(y)\subseteq M^{\lambda+}(y)$ by definition. \hfill
$\Box$

\vspace{1em}

We are now in the appropriate position to elaborate the difference
between the geometrical equivalence shown in this paper and the
asymptotical equivalence proven in \cite{DF05a}. The latter can be
expressed with our notations as $T^{\lambda-}(y)= M^{\lambda-}(y)$
while our result here indicates that
$\overline{T^{\lambda}(y)}=\overline{M^{\lambda}(y)}$. Since the
question whether or not $T^{\lambda-}(y)=\overline{T^{\lambda}(y)}$
(or equivalently, $M^{\lambda-}(y)=\overline{M^{\lambda}(y)}$)
remains open, our result cannot be derived directly from the one in
\cite{DF05a}.
\section{Conclusion}
To summarize, we show that in some cases catalyst-assisted
entanglement transformation is strictly more powerful than
multiple-copy one in either deterministic or probabilistic setting.
For purely probabilistic setting, however, we can prove that these
two kinds of transformations are geometrically equivalent in the
sense that the two sets $T^{\lambda}(y)$ and $M^{\lambda}(y)$,
denoting the sets of bipartite pure states which can be converted
into a given state with Schmidt coefficient vector $y$ with maximal
probabilities not less than $\lambda$ by catalyst-assisted
transformation and by multiple-copy transformation, respectively,
have the same closure. The limit properties of $T^{\lambda}(y)$ and
$M^{\lambda}(y)$ as set-valued functions about $\lambda$ are also
discussed.

The results about the relation between catalyst-assisted
transformation and multiple-copy transformation shown in this paper
and our previous works can be described by the following diagrams:
\begin{equation}
\begin{array}{ccccccccc}
M^{\lambda+}(y)&=&M^{\lambda}(y)^\circ&\varsubsetneq&M^{\lambda}(y)&\varsubsetneq&\overline{M^{\lambda}(y)}&\subsetq&
M^{\lambda-}(y)\\
\rotatebox{90}{=}&&\rotatebox{90}{=}&&\rotatebox{90}{$\varsupsetneq$}& &\rotatebox{90}{=}& & \rotatebox{90}{=}\\
T^{\lambda+}(y)&=&T^{\lambda}(y)^\circ&\varsubsetneq&T^{\lambda}(y)&\varsubsetneq&\overline{T^{\lambda}(y)}&\subsetq&
T^{\lambda-}(y)
\end{array}
\end{equation}
for purely probabilistic case ($\lambda<1$) and
\begin{equation}
\begin{array}{ccccccccc}
M(y)^\circ&\varsubsetneq&M(y)&\varsubsetneq&\overline{M(y)}&
\subsetq&
M^{-}(y)\\
\rotatebox{90}{\supsetq}&&\rotatebox{90}{$\varsupsetneq$}& &\rotatebox{90}{\supsetq}& & \rotatebox{90}{=}\\
T(y)^\circ&\varsubsetneq&T(y)&\varsubsetneq&\overline{T(y)}&\subsetq&
T^{-}(y)
\end{array}
\end{equation}
for deterministic case. Where we write $A(y)\varsubsetneq B(y)$ if
$A(y)\subseteq B(y)$ holds for all $y$ but there exists some $y$
such that $A(y)\neq B(y)$; while by $A(y)\subsetq B(y)$ we indicate
that whether or not there exists $y$ such that $A(y)\neq B(y)$ is
still open, although $A(y)\subseteq B(y)$ always holds for all $y$.

From the above two diagrams, the remaining questions for further
study are:

1). Whether or not $\overline{T^{\lambda}(y)}= T^{\lambda-}(y)$ (or
equivalently, $\overline{M^{\lambda}(y)}= M^{\lambda-}(y)$ ) for any
$y$ and $\lambda\leq 1$. In other words, whether or not the function
$T^{\lambda}(y)$ (or $M^{\lambda}(y)$) is `almost' left continuous
at any $\lambda\leq 1$.

2). Whether or not $\overline{T(y)}= \overline{M(y)}$ (or
equivalently, $T(y)^\circ= M(y)^\circ$) for any $y$. That is,
whether or not catalyst-assisted transformation and multiple-copy
transformation are also geometrically equivalent in deterministic
setting.

\section*{ACKNOWLEDGMENTS}

The authors thank the colleagues in the Quantum Computation and
Quantum Information Research Group for useful discussion. This work
was partly supported by the Natural Science Foundation of China
(Grant Nos. 60503001, 60321002, and 60305005), and by Tsinghua Basic
Research Foundation (Grant No. 052220204). R. Duan acknowledges the
financial support of Tsinghua University (Grant No. 052420003).

\bibliography{relation}

\end{document}